\documentclass[pra,reprint,twocolumn,superscriptaddress,showpacs,floatfix]{revtex4-1}
\usepackage{amsmath}
\usepackage{changes}
\usepackage{mathrsfs}
\usepackage{txfonts}
\usepackage{amssymb}
\usepackage{color}
\usepackage{graphicx}
\usepackage{hyperref}
\usepackage{ulem}
\usepackage{overpic}
\usepackage{psfrag}
\usepackage{tabularx}
\usepackage{array}
\usepackage{placeins}
\newcommand{\PreserveBackslash}[1]{\let\temp=\\#1\let\\=\temp}
\usepackage{color}
\makeatletter

\begin{document}
\title{Fractal dimension and the counting rule of the Goldstone modes}
\author{Qian-Qian Shi}
\affiliation{Centre for Modern Physics,
Chongqing University, Chongqing 400044, The People's Republic of
China}

\author{Yan-Wei Dai}
\affiliation{Centre for Modern Physics,
Chongqing University, Chongqing 400044, The People's Republic of
China}

\author{Huan-Qiang Zhou}\thanks{E-mail: hqzhou@cqu.edu.cn}
\affiliation{Centre for Modern Physics,
Chongqing University, Chongqing 400044, The People's Republic of China}

\author{Ian P. McCulloch}
\affiliation{ Department of Physics, National Tsing Hua University, Hsinchu 30013, Taiwan}
\affiliation{School of Mathematics and Physics, The University of Queensland, St. Lucia, QLD 4072, Australia}
\affiliation{Centre for Modern Physics, Chongqing University, Chongqing 400044, The People's Republic of China}

\begin{abstract}
 It is argued that  there are a set of orthonormal basis states, which appear  as highly degenerate ground states arising from spontaneous symmetry breaking with a type-B Goldstone mode,  and they are scale-invariant, with a salient feature that the entanglement entropy $S(n)$ scales logarithmically with the block size $n$ in the thermodynamic limit.  As it turns out, the prefactor is half the number of type-B Goldstone modes $N_B$.   This is achieved by performing an exact Schmidt decomposition of  the orthonormal basis states, thus unveiling their self-similarities  in the real space--the essence of a fractal.
Combining with a field-theoretic prediction [O. A. Castro-Alvaredo and B. Doyon, Phys. Rev. Lett. \textbf{108}, 120401 (2012)], we are led to the identification of the fractal dimension $d_f$ with the number of type-B Goldstone modes $N_B$ for the orthonormal basis states in quantum many-body systems undergoing spontaneous symmetry breaking.
 \end{abstract}
\maketitle

\section{introduction}
Spontaneous symmetry breaking (SSB) is a key ingredient in diverse areas of physics, ranging from condensed matter to field theories. In particular, the emergence of a gapless Goldstone mode (GM), when a continuous symmetry group is spontaneously broken, is of paramount importance, due to its relevance to the low-energy physics.  As first stated by Goldstone~\cite{goldstone}, the number of  GMs is equal to the number of broken symmetry generators $N_{BG}$ for a relativistic system undergoing SSB. However, complications arise for a  nonrelativistic system, as far as the connection between the number of broken symmetry generators $N_{BG}$ and the number of GMs is concerned. Since an early work by Nielsen and Chadha~\cite{nielsen}, much attention has been paid to a proper classification of GMs~\cite{schafer, miransky, nicolis, brauner-watanabe, watanabe, watanabe2, NG}, culminating in the introduction of type-A and type-B GMs~\cite{watanabe,watanabe2}, based on a previous observation made by Nambu~\cite{nambu}. In this classification, the so-called Watanabe-Brauner matrix~\cite{brauner-watanabe} plays a crucial role. As a result, when the symmetry group $G$ is spontaneously broken into $H$, the counting rule for GMs may be formulated as follows
 \begin{equation}
N_A+2N_B=N_{BG},\label{nab}
 \end{equation}
where $N_A$ and $N_B$ are, respectively, the numbers of type-A and type-B GMs, and $N_{BG}$ is equal to the dimension of the coset space $G/H$.

One remarkable distinction may be made between type-A and type-B GMs, since SSB with a type-A GM only happens in the thermodynamic limit, in contrast to SSB with a type-B GM, which survives in a finite-size system. Instead, a finite-size precursor to SSB with a type-A GM appears in the guise of the so-called Anderson tower, first developed in spin wave theory for antiferromagnetism~\cite{anderson}.
Meanwhile, a significant development has been achieved to describe SSB with a type-A GM from the perspective of quantum entanglement~\cite{Metlitski, Rademaker}.
However, a systematic investigation is still lacking for SSB with a type-B GM from the perspective of the entanglement entropy, with a few notable exceptions~\cite{ popkov1,salerno,doyon}, in which the entanglement entropy is discussed for the $\rm{SU(2)}$ Heisenberg ferromagnetic states.
The other distinction between type-A and type-B GMs concerns their instabilities under quantum fluctuations.
In fact, SSB with a type-A GM is forbidden in one spatial dimension, as a result of the Mermin-Wagner-Coleman theorem~\cite{mwc}, whereas SSB with a type-B GM survives quantum fluctuations even in one spatial dimension.
As a consequence, instead of long-range order resulted from SSB with a type-A GM, there exists only quasi-long-range order in one spatial dimension, which may be characterized by means of conformal field theory~\cite{cft}.
Historically, conformal field theory originated from a speculation made by Polyakov~\cite{polyakov} that scale invariance implies conformal invariance, which itself has attracted much attention, in an attempt to prove or disprove it~\cite{morecf,morecfsierra, morecf2}.
In this regard, an intriguing question arises as to whether or not there is any scale-invariant, but not conformally invariant state, if one takes into account SSB with a type-B GM, which survives even in one spatial dimension.

This work attempts to address this question through a thorough investigation of the scaling behavior of the entanglement entropy for one-dimensional quantum many-body systems undergoing SSB with a type-B GM.
We demonstrate that there are a set of orthonormal basis states, which appear  as highly degenerate ground states arising from SSB with a type-B GM, and they admit an exact Schmidt decomposition~\cite{nielsenchuang}, thus unveiling their self-similarities in the real space - the essence of a fractal, characterized in term of the fractal dimension $d_f$.  As a consequence,
highly degenerate ground states are scale-invariant, which in turn implies that the entanglement entropy $S(n)$ scales logarithmically with the block size $n$ in the thermodynamic limit.  As it turns out, the prefactor is half the number of type-B GMs $N_B$. Combining with a field-theoretic prediction~\cite{doyon} that the prefactor is half the fractal dimension $d_f$, we are led to the identification of the fractal dimension $d_f$ with the number of type-B GMs $N_B$  for the orthonormal basis states in quantum many-body systems undergoing SSB.  As an illustration, we investigate the $\rm{SU(2)}$ spin-$s$ ferromagnetic model, the $\rm{SU(N+1)}$ ferromagnetic model, with $N=2$, $3$ and $4$, and the $\rm{SU(2)}$ spin-1 anisotropic biquadratic model, with the fractal dimension $d_f$ being 1, $N$, and 1, respectively.

\section{The entanglement entropy for a type of scale-invariant states}

\subsection{An exact Schmidt decomposition and scale invariance}
Consider a translation-invariant quantum many-body system, described by the Hamiltonian $\mathscr{H}$, with the symmetry group $\rm{SU(N+1)}$, on a one-dimensional lattice. Throughout this work, the size $L$ is assumed to be even.
The symmetry group $\rm{SU(N+1)}$ has $(N+1)^2-1$ generators, and the rank is $N$.
Accordingly, there are $N$ commuting Cartan generators~\cite{gsin} $H_\alpha$ ($\alpha=1$, \ldots, $N$), which are traceless and diagonal.
For each $H_\alpha$, there exists a conjugate pair of a raising operator $E_\alpha$ and a lowering operator $F_\alpha$ such that $[H_\alpha, E_\beta]=\beta_\alpha E_\beta$, $[E_\alpha,F_\alpha]=(E_\alpha,F_\alpha)H_\alpha$, $[F_\alpha,H_\beta]=\beta_\alpha F_\beta$,  $[E_\alpha,E_\beta]=g_{\alpha,\;\beta}E_\gamma\delta_{\gamma,\;\alpha+\beta}$, and $[F_\alpha,F_\beta]=g_{-\alpha,\;-\beta}F_\gamma\delta_{\gamma,\;\alpha+\beta}$, with $\beta$ being the root matrix, $g_{\alpha,\;\beta}$ depending on the specific form of the Cartan generators, and $(E_\alpha,F_\alpha)$ being the Killing form of $E_\alpha$ and $F_\alpha$.
Here, we stress that, it is convenient to choose the Cartan generators $H_\alpha$ in such a way that the set of the lowering operators $F_\alpha$ commute with each other.

Suppose the symmetry group $\rm{SU(N+1)}$ is spontaneously broken into $\rm{SU(N)}\times\rm{U(1)}$.
For simplicity, we assume that the translational symmetry under the one-site translation is not spontaneously broken.
Otherwise, a unit cell is needed.  With this fact in mind, the highest weight state $|{\rm hws}\rangle$, which itself is an unentangled ground state, takes the form
$|{\rm hws}\rangle=|hh\cdots h\rangle$, with a local component $|h\rangle_j$ being the eigenvector for $H_{\alpha,j}$, satisfying $E_{\alpha,j}|h\rangle_j=0$, but $F_{\alpha,j}|h\rangle_j \neq 0$. Here, $H_{\alpha,j}$, $E_{\alpha,j}$, and $F_{\alpha,j}$ represent the local counterparts of the Cartan generators $H_\alpha$, the raising operators $E_\alpha$ and the lowering operators $F_\alpha$  at a lattice site $j$ on a one-dimensional lattice: $H_\alpha=\sum_jH_{\alpha,j}$, $E_\alpha=\sum_jE_{\alpha,j}$ and $F_\alpha=\sum_jF_{\alpha,j}$.
For the symmetry generators $E_\alpha$ and $F_\alpha$, one may choose $F_{\alpha,j}$ and $E_{\alpha,j}$ as the interpolating fields~\cite{nielsen, inter}, respectively. Given $\langle[E_\alpha,F_{\alpha,j}]\rangle\propto \langle H_{\alpha,j}\rangle$, $\langle[E_{\alpha,j},F_\alpha]\rangle\propto \langle H_{\alpha,j}\rangle$ and $\langle H_{\alpha,j}\rangle\neq0$, the $2N$ symmetry generators $E_\alpha$ and $F_\alpha$ are spontaneously broken, with $\langle H_{\alpha,j}\rangle$ being a local order parameter.
Here, the expectation value $\langle  O \rangle$ of an operator $O$ is taken over the highest weight state $|{\rm hws}\rangle$.
Since no type-A GM survives in one spatial dimension~\cite{mwc}, the number of type-A GMs $N_A$ must be 0. Therefore, according to the counting rule~(\ref{nab}), the $N$ type-B GMs emerge.
For a later use, we introduce $q_\alpha$ to denote the power of $F_{\alpha,j}$ such that $F_{\alpha,j}^{\,\,\,q_\alpha}|h\rangle_j\neq0$, but $F_{\alpha,j}^{\,\,\,q_\alpha+1}|h\rangle_j=0$.

A sequence of degenerate ground states $|L,M_1,\ldots,M_N\rangle$ are generated from the repeated action of the lowering operators $F_\alpha$ on the highest weight state $|\rm{hws}\rangle$:
 \begin{equation}
 |L,M_1,\ldots,M_N\rangle=\frac{1}{Z(L,M_1,\ldots,M_N)}\prod_{\alpha=1}^N F_\alpha^{\,\,M_\alpha}|\rm{hws}\rangle,
 \label{lmnr}
 \end{equation}
where $Z(L,M_1,\ldots,M_N)$ is introduced to ensure that $|L,M_1,\ldots,M_N\rangle$ is normalized.

In order to understand SSB with a type-B GM from the perspective of quantum entanglement, the system is partitioned into a block $\rm{B}$ and its environment $\rm{E}$.
Here, the block $\rm{B}$ consists of $n$ lattice sites that are not necessarily contiguous, with the rest $L-n$ lattice sites constituting the environment $\rm{E}$.
As a convention, $n\leq L/2$.
Note that $|{\rm hws}\rangle$, as an unentangled product state, is split into $|{\rm hws}\rangle_{\rm B}$ and $|{\rm hws}\rangle_{\rm E}$.
With this in mind, we introduce the counterparts of the symmetry group $\rm{SU(N+1)}$ in the block $\rm{B}$ and the environment $\rm{E}$, respectively.
In particular, the counterparts of the lowering operators $F_\alpha$ are $F_{\alpha,{\rm B}}$ and $F_{\alpha,{\rm E}}$, respectively, in the block $\rm{B}$ and the environment $\rm{E}$. Hence, we have $F_{\alpha}=F_{\alpha,{\rm B}}+ F_{\alpha,{\rm E}}$.
This in turn allows us to define the basis states $|n,k_1,\ldots,k_N\rangle$ and $|L-n,M_1-k_1,\ldots,M_N-k_N\rangle$ for the block $\rm{B}$ and the
environment $\rm{E}$, which take the same form as Eq.~(\ref{lmnr}), with $F_\alpha$ replaced by $F_{\alpha,{\rm B}}$ and $F_{\alpha,{\rm E}}$, $M_\alpha$ replaced by $k_\alpha$ and $M_\alpha-k_\alpha$, respectively.
Meanwhile, $Z(n,k_1,\ldots,k_N)$ and $Z(L-n,M_1-k_1,\ldots,M_N-k_N)$ need to be introduced to ensure that $|n,k_1,\ldots,k_N\rangle$ and $|L-n,M_1-k_1,\ldots,M_N-k_N\rangle$ are normalized.
\begin{widetext}	
A remarkable fact is that degenerate ground states $|L,M_1,\ldots,M_N\rangle$ admit an exact Schmidt decomposition:
	\begin{align}
		|L,M_1,\ldots,M_N\rangle=\sum_{k_1=0}^{\min(M_1,q_1 n)}\cdots \sum_{k_N=0}^{\min(M_N,q_N n)}\lambda(L,n,k_1,\ldots,k_N,M_1,\ldots,M_N)
		|n,k_1,\ldots,k_N\rangle\;|L-n,M_1-k_1,\ldots,M_N-k_N\rangle,
		\label{psignln}
	\end{align}
where the  Schmidt coefficients $\lambda(L,n,k_1,\ldots,k_N,M_1,\ldots,M_N)$ take the form
	\begin{align}
		\lambda(L,n,k_1,\ldots,k_N,M_1,\ldots,M_N)=C_{M_1}^{k_1}\cdots C_{M_N}^{k_N}
		\frac{Z(n,k_1,\ldots,k_N)Z(L-n,M_1-k_1,\ldots,M_N-k_N)}{Z(L,M_1,\ldots,M_N)}.
		\label{lambdak}
	\end{align}
Here $C_{M_\alpha}^{k_\alpha}$ denote the binomial coefficients: $C_{M_\alpha}^{k_\alpha}= M_\alpha !/ k_\alpha ! (M_\alpha - k_\alpha)!$ ($\alpha=1$, \ldots, $N$), originating from the binomial expansion: $(F_{\alpha,{\rm B}}+ F_{\alpha,{\rm E}})^{\,\,M_\alpha}=\sum_{k_\alpha=0}^{M_\alpha}C_{M_\alpha}^{k_\alpha}F_{\alpha,{\rm B}}^{\,\,k_\alpha}F_{\alpha,{\rm {\rm E}}}^{\,\,M_\alpha-k_\alpha}$. We remark that the three sets of quantum states $|L,M_1,\ldots,M_N\rangle$,
$|n,k_1,\ldots,k_N\rangle$ and $|L-n,M_1-k_1,\ldots,M_N-k_N\rangle$ constitute orthonormal basis states in the system, the block and the environment, respectively.	We remark that the three sets are generated from the action of lowering operators (of the same group) on the highest weight states, defined for the block, the environment and the system, respectively.  That is, they are similar
to each other, in the sense that they are identical after performing a scale transformation connecting the block, the environment and the system.  Mathematically,  the self-similarities in the real space are embodied in the fact 
that, if a scale transformation $n\leftrightarrow L$ is performed, then $|n,k_1,\ldots,k_N\rangle \leftrightarrow |L,M_1,\ldots,M_N\rangle$,  $|{\rm hws}\rangle_{\rm B}\leftrightarrow |\rm{hws}\rangle$, 
$F_{\alpha,{\rm B}}^{\,\,k_\alpha} \leftrightarrow F_{\alpha}^{\,\,M_\alpha}$ ($\alpha=1$, \ldots, $N$) and $Z(n,k_1,\ldots,k_N)\leftrightarrow Z(L,M_1,\ldots,M_N)$.
This requires that the numbers of the orthonormal basis states for both the entire system and the subsystems must match, if a proper scale transformation is taken into account. 
Hence, the self-similarity manifests itself in the exact Schmidt decomposition (\ref{psignln}).
In other words, the Schmidt decomposition  (\ref{psignln}) reflects the self-similarities in the real space -- the essence of an intrinsic abstract fractal underlying the ground state subspace. This explains why the fractal dimension $d_f$,  as already exploited in a field-theoretic approach to the  ${\rm SU}(2)$ ferromagnetic states~\cite{doyon}, furnishes a proper description for this specific type of scale-invariant state.

The entanglement entropy $S_L(n,M_1,\ldots,M_N)$ follows from the reduced density matrix $\rho_L(n,M_1,\ldots,M_N)$:
$S_L(n,M_1,\ldots,M_N)=-{\rm Tr}[\rho_L(n,M_1,\ldots,M_N)\log_2\rho_L(n,M_1,\ldots,M_N)]$, with
\begin{equation}
\rho_L(n,M_1,\ldots,M_N)=\sum_{k_1=0}^{\min(M_1,q_1 n)}\cdots \sum_{k_N=0}^{\min(M_N,q_N n)}\Lambda(L,n,k_1,\ldots,k_N,M_1,\ldots,M_N)
|n,k_1,\ldots,k_N\rangle\langle n,k_1,\ldots,k_N|.
\end{equation}
Here, $\Lambda(L,n,k_1,\ldots,k_N,M_1,\ldots,M_N)$ are the eigenvalues of the reduced density matrix $\rho_L(n,M_1,\ldots,M_N)$: $\Lambda(L,n,k_1,\ldots,k_N,M_1,\ldots,M_N)=[\lambda(L,n,k_1,\ldots,k_N,M_1,\ldots,M_N)]^2$.
Therefore, $S_L(n,M_1,\ldots,M_N)$ may be rewritten as
\begin{align}
S_L(n,M_1,\ldots,M_N)=-\sum_{k_1=0}^{\min(M_1,q_1 n)}\cdots \sum_{k_N=0}^{\min(M_N,q_N n)} \Lambda(L,n,k_1,\ldots,k_N,M_1,\ldots,M_N)
\log_{2}\Lambda(L,n,k_1,\ldots,k_N,M_1,\ldots,M_N).
\label{snk}
\end{align}
\end{widetext}
This makes it possible to perform a systematic analysis of the block entanglement entropy $S_L(n,M_1,\ldots,M_N)$, depending on a specific realization of the symmetry group $\rm{SU(N+1)}$ for a concrete model Hamiltonian $\mathscr{H}$ under investigation. A detailed investigation will be carried out for the $\rm{SU(2)}$ spin-$s$ ferromagnetic model, the $\rm{SU(N+1)}$ ferromagnetic model, and the $\rm{SU(2)}$ spin-1 anisotropic biquadratic model in Sections~~\ref{fm1}, \ref{fm2} and \ref{fm3}.

\subsection{Scaling of the entanglement entropy with the block size in the thermodynamic limit}

Instead of directly investigating into the block entanglement entropy $S_L(n,M_1,\ldots,M_N)$ for a specific quantum many-body system, we turn to a scaling analysis of the entanglement entropy $S_L(n,M_1,\ldots,M_N)$ with the block size $n$ for degenerate ground states arising from SSB with type-B GMs in the thermodynamic limit, i.e., when $L \rightarrow \infty$.

For this purpose, we introduce the fillings $f_\alpha=M_\alpha/L$ ($\alpha=1,$ \ldots, $N$) to ensure that $f_\alpha$ are kept constant, when $L$
tends to infinity. To ease the notations, we denote $S(n)$ as the block entanglement entropy for a specific choice of the fillings $f_\alpha$ ($\alpha=1,$ \ldots, $N$). Physically, that amounts to restricting to  the orthonormal basis states (\ref{lmnr}), since they are the only degenerate ground states with well-defined values of the fillings  $f_\alpha=M_\alpha/L$.
Given that the self-similarity in the real space manifests itself in an intrinsic abstract fractal underlying the ground state subspace,  the orthonormal basis states, defined in Eq.(\ref{lmnr}), are scale-invariant.
With this observation in mind, we perform a scale transformation $n \rightarrow \lambda \; n$ that amounts to introducing two sequences of the values of the block size $n$, with a dimensionless ratio $\lambda$. Indeed, the scale invariance imposes the constraint on the entanglement entropy $S(n)$, thus leading to the establishment of a logarithmic scaling relation between the entanglement entropy $S(n)$ and the block size $n$ ($1\ll n \ll L/2$) (for a detailed derivation, cf. Sec. A of the Supplemental Material (SM)). More precisely, the scaling relation for the orthonormal basis states (\ref{lmnr}) takes an asymptotic form
\begin{equation}
 S(n)=\frac{N_B}{2}\log_2 n+S_0,
 \label{sn}
 \end{equation}
where $S_0$ is an additive contribution to the entanglement entropy, which depends on the fillings $f_\alpha$ ($\alpha=1,$ \ldots, $N$). Here, Eq.(\ref{sn}) should be understood as a scaling relation that is valid for sufficiently large $n$ in the thermodynamic limit, under the condition  that the fillings $f_\alpha$ ($\alpha=1,$ \ldots, $N$) are non-zero or the maximum. That is, it is necessary to exclude both the highest and lowest weight states, since they are generically product states, with the entanglement entropy being zero. Physically,  this is due to the fact that  both the highest and lowest weight states  do not accommodate type-B GMs, in contrast to other degenerate ground states.
Meanwhile, we emphasize that the prefactor in front of $\log_2 n$ is half the number of type-B GMs $N_B$ {\it only} for  the orthonormal basis states (\ref{lmnr}), it thus is universal, in the sense that it is model-independent. In fact, this follows from a simple physical consideration. Indeed, the prefactor, denoted as $\kappa$, must be a function of $N_B$: $\kappa = \kappa(N_B)$, since we focus on low energy physics. Imagine a fictitious system consisting of two arbitrary (interacting) subsystems that are not coupled to each other, with the numbers of  type-B GMs being $ N_{B,1}$ and $N_{B,2}$, respectively. Then, the total number of type-B GMs for the fictitious system is $N_B= N_{B,1}+N_{B,2}$. Accordingly, we have  $\kappa(N_B)=\kappa(N_{B,1})+\kappa(N_{B,2})$, given that the  entanglement entropy for the fictitious system is the sum of the counterparts for the two subsystems. This implies that $\kappa(N_B)$ is linearly proportional to $N_B$. We stress that the proportionality constant is a bit subtle for a generic degenerate ground state, which usually is a linear combination of the orthonormal basis states (\ref{lmnr}). However,  it may be determined to be $1/2$  for the orthonormal basis states (\ref{lmnr}), as long as  both the highest and lowest weight states are excluded. This follows from an exact asymptotic analysis for the $\rm{SU(2)}$ spin-$1/2$ Heisenberg ferromagnetic states~\cite{salerno}(also cf. Sec. B of the SM).

In addition, for the $\rm{SU(2)}$ Heisenberg ferromagnetic model, a scaling relation of the entanglement entropy $S(n)$ with the block size $n$ in the thermodynamic limit has been investigated for a linear combination of the overcomplete (non-orthogonal) basis states as a subset of the coset space~\cite{doyon}. As a result, $S(n)$ scales logarithmically with $n$, with the prefactor being half the fractal dimension $d_f$.
Hence, combining with this field-theoretic prediction~\cite{doyon}, we conclude that the fractal dimension $d_f$ is identical to the number of type-B GMs $N_B$ for  the orthonormal basis states (\ref{lmnr}) in the ground state subspace.
Here we remark that such a logarithmic scaling relation is valid for any state in the ground state subspace of a quantum many-body system undergoing SSB with type-B GMs, which may be expressed  in terms of a linear combination of  the orthonormal basis states (\ref{lmnr}), with the prefactor being relevant to the number of type-B GMs.  Generically, the relation between the prefactor $\kappa$ and the number of type-B GMs $N_B$ is a bit involved. In fact, it is necessary to introduce an extrinsic fractal, e.g., the Cantor sets, as  a support to express this degenerate ground state in terms of a linear combination of the overcomplete (non-orthogonal) basis states. If the same type of a Cantor set is adopted for each of the type-B GMs, then the prefactor $\kappa$ is proportional to the number of type-B GMs.  In particular, the proportionality constant is $1/2$ only for the orthonormal basis states generated from the action of the lowering operators on the highest weight state, since the fractal dimension of the support is identical to the number of type-B GMs in this particular case  (more detailed discussions may be found in a forthcoming article~\cite{cantorset}). Here we stress that the orthonormal basis states considered here  constitute {\it only} a subset of the set of all possible linear combinations of the overcomplete basis states. A notable feature is that the orthonormal basis states are nothing but highly degenerate ground states admitting a natural interpretation in the group representation-theoretic context, given that they are constructed in a model-independent way, valid for any semisimple symmetry group, though the orthonormal basis states themselves are model-dependent.

The scaling relation (\ref{sn}) between the block entanglement entropy $S(n)$ and the block size $n$ constitutes the main result of this work. The remaining task is to investigate the block entanglement entropy $S_L(n,M_1,\ldots,M_N)$ for three illustrative quantum many-body systems and demonstrate how such a logarithmic scaling relation emerges from numerical calculations, as the system size $L$ tends to infinity.

\section{The $\rm{SU(2)}$ ferromagnetic states: arbitrary spin $s$}~\label{fm1}
Consider the $\rm{SU(2)}$ spin-$s$ ferromagnetic Heisenberg model with the nearest-neighbor interaction, described by the Hamiltonian
\begin{equation}
\mathscr{H}=-\sum_{j=1}^{L}\textbf{S}_j\cdot \textbf{S}_{j+1}, \label{Hsu2}
\end{equation}
where $\textbf{S}_j=(S_{x,j},S_{y,j},S_{z,j})$, and $S_{x,j}$, $S_{y,j}$ and $S_{z,j}$ represent the spin-$s$ operators at the $j$-th site.
Here, the symmetry group $\rm{SU(2)}$ is generated by $S_+=\sum_jS_{+,j}$, $S_-=\sum_jS_{-,j}$ and $S_z=\sum_jS_{z,j}$: $[S_z,S_+]=S_+$, $[S_+,S_-]=S_z$ and $[S_-,S_z]=S_-$, with $S_{+,j}$ and $S_{-,j}$ being defined by $ S_{\pm,j}=(S_{x,j}\pm iS_{y,j})/\sqrt{2}$.
Suppose $|m\rangle_j$ are the eigenvectors of $S_{z,j}$: $S_{z,j}|m\rangle_j=m|m\rangle_j$, with $m=-s$, \ldots, $s$.
Then, the local Hilbert space constitutes a $2s+1$-dimensional irreducible representation of $\rm{SU(2)}$ at each lattice site $j$.
The action of $S_{-,j}$ and $S_{+,j}$ on $|m\rangle_j$ takes the form: $S_{-,j}|m\rangle_j=\sqrt{(s+m)(s-m+1)/2}|m-1\rangle_j$ and $S_{+,j}|m\rangle_j=\sqrt{(s-m)(s+m+1)/2}|m+1\rangle_j$, respectively.
Thus, the highest weight state $|{\rm hws}\rangle$ is $|{\rm hws}\rangle=|ss\cdots s\rangle$.
Since $S_{-}^{\;\;\;2s}|s\rangle\neq0$, but $S_{-}^{\;\;2s+1}|s\rangle=0$, we have $q=2s$.
Note that the model is exactly solvable by means of the Bethe ansatz only when $s=1/2$.

The interpolating fields are $S_{+,j}$ and $S_{-,j}$, for the generator $S_{-}$ and $S_{+}$, respectively.
Thus, $\langle S_{z,j}\rangle$ is the local order parameter, given $\langle[S_{+,j},S_{-}]\rangle=\langle[S_{+},S_{-,j}]\rangle=\langle S_{z,j}\rangle\neq0$.
Hence, the two generators $S_-$ and $S_+$ are spontaneously broken.
According to the counting rule~(\ref{nab}), the number of type-B GM is one.
Therefore, the model is a specific realization of the general scheme: $H_1=S_z$, $E_1=S_+$ and $F_1=S_-$.

The degenerate ground states $|L,M\rangle$ are generated from the repeated action of the lowering operator $S_-$ on the highest weight state $|ss \cdots s\rangle$:
\begin{equation}
|L,M\rangle=\frac{1}{Z(L,M)}S_-^{\,\,M}|ss\cdots s\rangle,
\end{equation}
where
\begin{equation}
Z(L,M) =\frac{M!}{\sqrt{2^M}}\!\sqrt{{\sum}'_{N_{-s},\ldots,\;N_s}\!\!\prod_{r=-s}^{s-1}\left[\varepsilon(s,r)\right]^{N_{r}}
{C_{L\!-\!\sum_{m=-s}^{r-1}N_m}^{N_r}}}\;,
\label{zlmsu2}
\end{equation}
with
\begin{equation}
\varepsilon(s,r)=\frac{\prod_{m=r+1}^{s}{(s+m)(s-m+1)}}{\prod_{m=r}^{s-1}(s-m)^2}.
\end{equation}
Here, $\sum'_{N_{-s},\ldots,\;N_s}$ is taken over all the possible values of $N_{-s}$, \ldots, $N_s$, subject to the constraints: $\sum_{m=-s}^s N_m=L$ and $\sum_{m=-s}^{s}(s-m)N_m=M$.
We remark that $|L,M\rangle$ ($M=0,$ \ldots, $2sL$) span a $2sL$+1-dimensional irreducible representation of the symmetry group $\rm{SU(2)}$.
A derivation of the concrete expression for $Z(L,M)$ is presented in Sec. C of the SM.

The degenerate ground states $|L,M\rangle$ admit an exact  Schmidt decomposition:
\begin{equation}
|L,M\rangle=\sum_{k=0}^{\min{(M,2sn)}}\lambda(L,n,k,M)|n,k\rangle|L-n,M-k\rangle,
\end{equation}
where the Schmidt coefficients $\lambda(L,n,k,M)$ take the form
\begin{equation}
\lambda(L,n,k,M)=\frac{\mu(L,n,k,M)}{\nu(L,n,k,M)},
\end{equation}
with
\begin{equation}
\mu(L,n,k,M)\!=\!\sqrt{\!{\sum}'_{n_{-\!s},\ldots,\; n_{s},\atop l_{-\!s},\ldots,\;l_{s}}\!\prod_{r,t=-s}^{s-1}\!\varepsilon(s,r)^{n_{r}}
{C_{n\!-\!\sum_{m=-s}^{r-1}\!n_m}^{n_r}}\!\varepsilon(s,t)^{l_{t}}{C_{L\!-\!n-\!\sum_{m=-s}^{t-1}l_m}^{l_t}}},
\end{equation}
and
\begin{equation}
\nu(L,n,k,M)=\sqrt{{\sum}'_{N_{-s}, \ldots,N_{s}}\prod_{r=-s}^{s-1}\varepsilon(s,r)^{N_{r}}
{C_{L-\sum_{m=-s}^{r-1}N_m}^{N_r}}}.
\end{equation}
Here, $\sum'_{n_{-s},\ldots,\;n_{s}}$ is taken over all the possible values of $n_{-s}$, \ldots, $n_s$, subject to the constraints: $\sum_{m=-s}^s n_m=n$ and $\sum_{m=-s}^{s}(s-m)n_m=k$, and $\sum'_{l_{-s},\ldots,\;l_{s}}$ is taken over all the possible values of $l_{-s}$, \ldots, $l_s$, subject to the constraints: $\sum_{m=-s}^s l_m=L-n$ and $\sum_{m=-s}^{s}(s-m)l_m=M-k$.
Then, the eigenvalues $\Lambda(L,n,M,k)$ of the reduced density matrix $\rho_L(n,M)$ follows from $\Lambda(L,n,M,k)=[\lambda(L,n,M,k)]^2$.
Note that the same results for spin $s=1/2$, presented in Ref.~\cite{popkov1}, are reproduced.
Hence, the entanglement entropy $S_L(n,M)$ follows from Eq.~(\ref{snk}).
In particular, the logarithmic scaling behaviour may be confirmed from an analytical treatment, based on the Stirling's approximation~\cite{stirling}, as done in Ref.~\cite{popkov1} for spin $s=1/2$ (also cf. Sec. B of the SM).

\begin{figure}[h]
	\centering
	\includegraphics[height=4.5cm,width=0.33\textwidth]{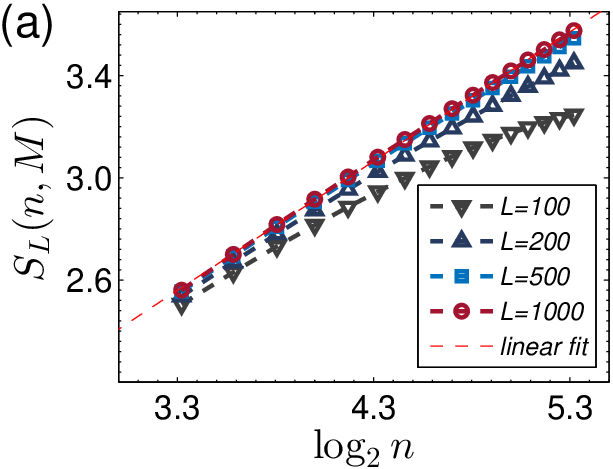}
	\includegraphics[height=4.5cm,width=0.33\textwidth]{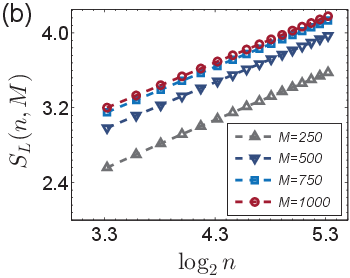}
	\includegraphics[height=4.5cm,width=0.33\textwidth]{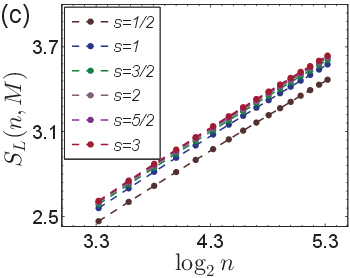}
	\caption{The entanglement entropy $S_L(n,M)$ vs $\log_2 n$ for the spin-$s$ $\rm{SU(2)}$ ferromagnetic states:
		(a) For $s=1$, $M=L/4$, when $L$ is varied: $L=$ 100, 200, 500 and
		1000. A significant deviation from the logarithmic scaling behaviour is observed when $L$ is relatively small, but tends to vanish, as $L$ increases.
		The prefactor is close to the exact value $1/2$, with an error being less than 1.1$\%$, when $L=1000$: $S_{1000}(n,250) = 0.505\log_2n + 0.891$.
		(b) For $s=1$, $L=1000$, when $M$ is varied: $M=$ $250$, $500$, $750$ and $1000$.
		The prefactor is close to $1/2$, within an error less than 2.3$\%$.
		(c) For $M=250$, $L=1000$, when $s$ is varied: $s=$ $1/2$, 1, 3/2, 2, 5/2 and $3$.
		The prefactor is close to $1/2$ for any spin $s$, within an error less than 2.2$\%$.
		Here, $n$ ranges from $10$ to $40$.
	}
	\label{Su2many}
\end{figure}

In order to understand how the logarithmic scaling behaviour emerges in the thermodynamic limit, we plot $S_L(n,M)$ vs $\log_2 n$ in Fig.~\ref{Su2many}:  (a) For $s=1$, $M=L/4$, when $L$ is varied: $L=$ 100, 200, 500 and 1000.
A significant deviation from the logarithmic scaling behaviour is observed when $L$ is relatively small, but tends to vanish, as $L$ increases.
Indeed, the prefactor is close to the exact value $1/2$, with an error being less than 1.1$\%$, when $L=1000$: $S_{1000}(n,250) = 0.505\log_2n + 0.891$.
(b) For $s=1$, $L=1000$, when $M$ is varied: $M=$ $250$, $500$, $750$ and $1000$.
This amounts to varying the filling $f$.
The prefactor is close to $1/2$, regardless of the values of the filling $f$, within an error less than 2.3$\%$.
That is, the contribution from the filling $f$ goes to a non-universal additive constant, as anticipated.
(c) For $M=250$, $L=1000$, when $s$ is varied: $s=$ $1/2$, 1, 3/2, 2, 5/2 and $3$.
The prefactor is close to $1/2$ for any spin $s$, within an error less than 2.2$\%$.
Here, $n$ ranges from $10$ to $40$.
Our numerics confirm that, for large enough $L$,  $S_L(n,M)$ logarithmically increases with $n$, as $n$ increases, subject to $1\ll n \ll L/2$, irrespective of $M$, as long as $M/L$ is neither 0 nor 1. That is, for the block size $n$ between 10 and 40,  it is necessary to choose, e.g., $L=1000$,  to ensure that it is large enough to reach a scaling limit.

\section{The $\rm{SU(N+1)}$ ferromagnetic states: fundamental representation}~\label{fm2}

The $\rm{SU(N+1)}$ ferromagnetic model is described by the Hamiltonian
\begin{equation}
\mathscr{H}=-\sum_{j}P_{j\;j+1}. \label{HsuNp1}
\end{equation}
Here, $P$ is the permutation operator:
$P=\sum_{u,v=1}^{N+1} e_{uv}\otimes e_{vu}$,
where
$e_{uv}=|u\rangle\langle v|$, with $|u\rangle$ and $|v\rangle$ being the $u$-th and $v$-th states in an orthonormal basis.
Physically, the permutation operator $P$ may be realized in terms of the spin-$s$ operators $\textbf{S}=(S_x,S_y,S_z)$, with $N=2s$:
\begin{equation*}
P=\sum_{r=0}^{2s}(-1)^{2s+r} \prod_{m\neq r}^{2s} \frac{2(\textbf{S}\otimes\textbf{S})-m(m+1)+2s(s+1)}{r(r+1)-m(m+1)}.
\end{equation*}
Note that, when $N=2$, it is the $\rm{SU(3)}$ ferromagnetic point for the spin-$1$ bilinear-biquadratic model~\cite{cfJxyz}.  The model is exactly solvable by means of the Bethe ansatz~\cite{sutherland}.

The model possesses the symmetry group $\rm{SU(N+1)}$, with the local Hilbert space being the fundamental representation space of $\rm{SU(N+1)}$ at each lattice site $j$, thus constituting a specific realization of the general scheme.
The Cartan generators $H_\alpha=\sum_jH_{\alpha,j}$ may be chosen as $H_{\alpha,j}=e_{11,j}-e_{\alpha+1\;\alpha+1,j}$ for $\alpha=1$, \ldots, $N$.
For each $H_{\alpha}$, the lowering operator and the raising operator may be chosen as: $F_\alpha=\sum_jF_{\alpha,j}$ $E_\alpha=\sum_jE_{\alpha,j}$, with $F_{\alpha,j}=e_{\alpha+1\;1,j}$ and $E_{\alpha,j}=e_{1\;\alpha+1,j}$, satisfying
$[H_\alpha,E_\alpha]=2E_\alpha$, $[E_\alpha,F_\alpha]=H_\alpha$ and $[F_\alpha,H_\alpha]=2F_\alpha$.
Define $|\beta\rangle$ as a $N+1$-dimensional vector, with the $\beta$-th entry being 1 and the others being 0.
Then, $|\beta\rangle_j$ are the eigenvectors of $H_{\alpha,j}$:
$H_{\alpha,j}|\beta\rangle_j=(\delta_{1\;\beta}-\delta_{\alpha+1\;\beta})|\beta\rangle_j$, for $\beta=1$, \ldots, $N+1$.
The action of $F_{\alpha,j}$ and $E_{\alpha,j}$ on $|1\rangle_j$ takes the form: $F_{\alpha,j}|1\rangle_j=|\alpha+1\rangle_j$ and $E_{\alpha,j}|1\rangle_j=0$.
Therefore, the highest weight state $|{\rm hws}\rangle$ is $|{\rm hws}\rangle=|11 \cdots1\rangle$.
The interpolating fields are $E_{\alpha,j}$ and $F_{\alpha,j}$, for $F_{\alpha}$ and $E_{\alpha}$, respectively.
Thus, $\langle H_{\alpha,j}\rangle$ ($\alpha=1$, \ldots, $N$) are the local order parameters, given $\langle[E_{\alpha,j},F_\alpha]\rangle=\langle[E_{\alpha},F_{\alpha,j}]\rangle=\langle H_{\alpha,j}\rangle\neq0$.
Hence, the $2N$ symmetry generators $E_\alpha$ and $F_\alpha$ are spontaneously broken.
According to the counting rule~(\ref{nab}), the $2N$ broken generators yield $N$ type-B GMs.
In addition, since $F_{\alpha,j}|1\rangle_j\neq0$, but $F_{\alpha,j}^2|1\rangle_j=0$, we have $q_\alpha=1$ ($\alpha=1$, \ldots, $N$).

 The degenerate ground states $|L,M_1,\ldots,M_N\rangle$ are generated from the repeated action of the lowering operators $F_{\alpha}$ on the highest weight state $|11\cdots1\rangle$:
\begin{equation}
|L,M_1,\ldots,M_N\rangle
=\frac{1}{Z(L,M_1,\ldots,M_N)}\prod_{\alpha=1}^{N}F_\alpha^{\,\,M_\alpha}|11\cdots 1\rangle,
\label{gsineq}
\end{equation}
which span an irreducible representation of the symmetry group $\rm{SU(N+1)}$, with the dimension being $C_{L+N}^N$.
Here, $Z(L,M_1,\ldots,M_N)$ takes the form,
\begin{equation}
Z(L,M_1,\ldots,M_N)=\prod_{\alpha=1}^N M_\alpha!\sqrt{C_{L-\sum_{\beta=1}^{\alpha-1}{M_\beta}}^{M_{\alpha}}} \;.
\label{zlmsuN}
\end{equation}
A derivation of the concrete expression for $Z(L,M_1,\ldots,M_N)$ is presented in Sec. D of the SM.
\begin{widetext}
The degenerate ground states $|L,M_1,\ldots,M_N\rangle$ admit an exact  Schmidt decomposition:
\begin{align}
|L,M_1,\ldots,M_N\rangle=
\sum_{k_1=0}^{\min(M_1,n)}\cdots \sum_{k_N=0}^{\min(M_N, n)}\lambda(L,n,k_1,\ldots,k_N,M_1,\ldots,M_N)
|n,k_1,\ldots,k_N\rangle|L-n,\!M_1-k_1,\ldots,M_N-k_N\rangle,
\end{align}
where the Schmidt coefficients $\lambda(L,n,k_1,\ldots,k_N,M_1,\ldots,M_N)$ take the form
\begin{align}
\lambda(L,n,k_1,\ldots,k_N,M_1,\ldots,M_N)=
\sqrt{\frac{\prod_{\alpha=1}^N {C_{n-\sum_{\beta=1}^{\alpha-1}{k_\beta}}^{k_{\alpha}}\prod_{\gamma=1}^N C_{L-n-\sum_{\beta=1}^{\gamma-1}{(M_\beta-k_\beta)}}^{M_\gamma-k_{\gamma}}}}
{\prod_{\alpha=1}^N {C_{L-\sum_{\beta=1}^{\alpha-1}{M_\beta}}^{M_{\alpha}}}}} \;.
\end{align}

Therefore, the eigenvalues $\Lambda(L,n,k_1,\ldots,k_N,M_1,\ldots,M_N)$ of the reduced density matrix $\rho_L(n,M_1,\ldots,M_N)$ are $\Lambda(L,n,k_1,\ldots,k_N,M_1,\ldots,M_N)=[\lambda(L,n,k_1,\ldots,k_N,M_1,\ldots,M_N)]^2$.
Hence, the entanglement entropy $S_L(n,M_1,\ldots,M_N)$ follows from Eq.~(\ref{snk}).  Thus, we reproduce the eigenvalues of the reduced density matrix $\rho_L(n,M_1,\ldots,M_N)$ in Ref.~\cite{salerno}.
In particular, the logarithmic scaling behaviour for the entanglement entropy $S_L(n,M_1,\ldots,M_N)$  may be confirmed from an analytical treatment~\cite{salerno}(also cf. Sec. B of the SM).
\end{widetext}

\begin{figure}
	\centering
	\includegraphics[height=4.5cm,width=0.33\textwidth]{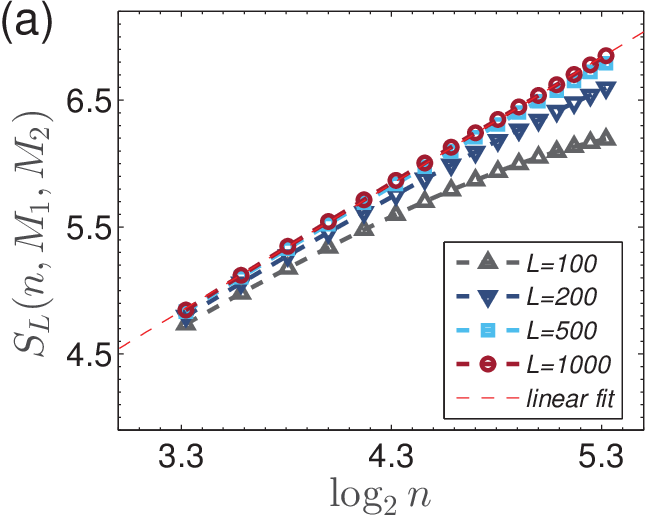}
	\hspace{5mm}
	\includegraphics[height=4.5cm,width=0.33\textwidth]{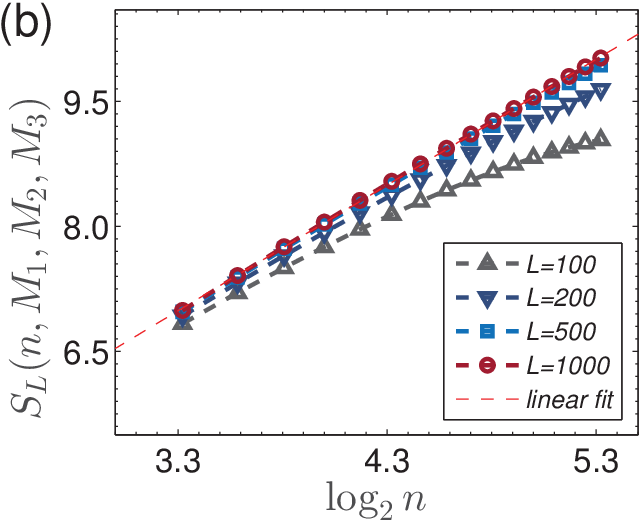}
	\hspace{5mm}
	\includegraphics[height=4.5cm,width=0.33\textwidth]{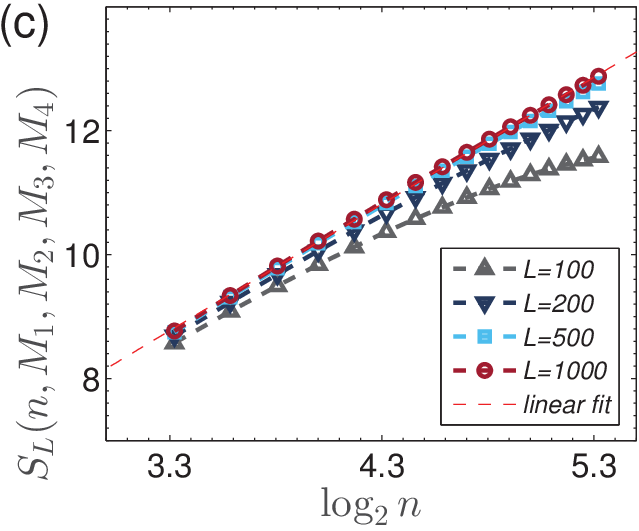}
	\caption{(a) The entanglement entropy $S_L(n,M_1,M_2)$ vs $\log_2 n$ for the $\rm{SU(3)}$ ferromagnetic states, with $M_1=L/2$ and $M_2=L/4$.
		(b) The entanglement entropy $S_L(n,M_1,M_2,M_3)$ vs $\log_2 n$ for the $\rm{SU(4)}$ ferromagnetic states, with $M_1=M_2=M_3=L/4$.
		(c) The entanglement entropy $S_L(n,M_1,M_2,M_3,M_4)$ vs $\log_2 n$ for the $\rm{SU(5)}$ ferromagnetic states, with $M_1=M_2=M_3=M_4=L/5$. Here, $L$ is varied: $L=100$, $200$, $500$ and $1000$.
		A significant deviation from the logarithmic scaling behaviour is observed when $L$ is relatively small, but tends to vanish, as $L$ increases.
		The prefactor is close to the exact value $N_B/2$, with an error being less than 2.3$\%$, when $L=1000$: $S_{1000}(n,500,250) = 0.999\log_2n + 1.542$, $S_{1000}(n,250,250,250) = 1.509\log_2n + 2.007$ and $S_{1000}(n,200,200,200,200) = 2.045\log_2n + 2.023$, with  $N_B$ being $2$, $3$ and $4$, respectively,  for the $\rm{SU(3)}$, $\rm{SU(4)}$ and $\rm{SU(5)}$ ferromagnetic states.
		Here, $n$ ranges from $10$ to $40$.}
	\label{entropysuN}
\end{figure}

In order to understand how the logarithmic scaling behaviour emerges in the thermodynamic limit, we plot $S_L(n,M_1,M_2)$ vs $\log_2 n$ in Fig.~\ref{entropysuN} (a) for the $\rm{SU(3)}$ ferromagnetic states, with $M_1=L/2$ and $M_2=L/4$, $S_L(n,M_1,M_2,M_3)$ vs $\log_2 n$ in Fig.~\ref{entropysuN} (b) for the $\rm{SU(4)}$ ferromagnetic states, with $M_1=M_2=M_3=L/4$, and $S_L(n,M_1,M_2,M_3,M_4)$ vs $\log_2 n$ in Fig.~\ref{entropysuN} (c) for the $\rm{SU(5)}$ ferromagnetic states, with $M_1=M_2=M_3=M_4=L/5$, when $L$ is varied: $L=100$, $200$, $500$ and $1000$.
A significant deviation from the logarithmic scaling behaviour is observed when $L$ is relatively small, but tends to vanish, as $L$ increases.
The prefactor is close to the exact value $N_B/2$, with an error being less than 2.3$\%$, when $L=1000$: $S_{1000}(n,500,250) = 0.999\log_2n + 1.542$, $S_{1000}(n,250,250,250) = 1.509\log_2n + 2.007$ and $S_{1000}(n,200,200,200,200) = 2.045\log_2n + 2.023$,  with $N_B$ being $2$, $3$ and $4$, respectively,  for the $\rm{SU(3)}$, $\rm{SU(4)}$ and $\rm{SU(5)}$ ferromagnetic states.
Here, $n$ ranges from $10$ to $40$.
Our numerics confirm that, for large enough $L$,  $S_L(n,M_1,M_2)$, $S_L(n,M_1,M_2,M_3)$ and $S_L(n,M_1,M_2,M_3,M_4)$  logarithmically increases with $n$, as $n$ increases, subject to $1\ll n \ll L/2$. That is, for the block size $n$ between 10 and 40,  it is necessary to choose, e.g., $L=1000$,  to ensure that it is large enough to reach a scaling limit.

\section{The coexisting fractal states: an example beyond simple ferromagnetism}~\label{fm3}
Consider the $\rm{SU(2)}$ spin-1 anisotropic biquadratic model~\cite{cfJxyz}, described by the Hamiltonian
\begin{equation}
\mathscr{H}=\sum_{j}{(J_xS_{x,j}S_{x,j+1}+J_yS_{y,j}S_{y,j+1}+J_zS_{z,j}S_{z,j+1})^2}, \label{xxz}
\end{equation}
where $S_{x,j}$, $S_{y,j}$, and $S_{z,j}$ are the spin-1 operators at a lattice site $j$, and $J_x$,  $J_y$ and $J_z$ are the anisotropic coupling parameters. Here, we focus on the  model on a characteristic line $J_x=J_y$. In this particular case,
the model possesses the staggered symmetry group $\rm{SU(2)}$ generated by $K_x$, $K_y$, and $K_z$:
$K_x=\sum_jK_{x,j}$, $K_y=\sum_jK_{y,j}$ and $K_z=\sum_jK_{z,j}$, with $K_{x,j}=\sum_j(-1)^j[{S_{x,j}}^2-{S_{y,j}}^2]/2$, $K_{y,j}=\sum_j(-1)^j(S_{x,j}S_{y,j}+S_{y,j}S_{x,j})/2$ and $K_{z,j}=\sum_{j}S_{z,j}/2$ (for the details, we refer to Ref.~\cite{cfJxyz}).
Accordingly, one may define the raising operator $K_+=\sum_j{K_{+,j}}$ and the lowering operator $K_-=\sum_j{K_{-,j}}$, with $K_{\pm,j}=(K_{x,j}\pm iK_{y,j})/\sqrt{2}$: $[K_z,K_+]=K_+$, $[K_+,K_-]=K_z$ and $[K_-,K_z]=K_-$.
In addition, it enjoys two extra $\rm{U(1)}$ symmetry groups, generated by $K_1$ and $K_2$: $K_1=\sum_j (-1)^{j} [{S_{y,j}}^2-{S_{z,j}}^2]/2$ and $K_2=\sum_j (-1)^{j} [{S_{z,j}}^2-{S_{x,j}}^2]/2$, respectively.
Since $K_1+K_2+K_x=0$, we only need to consider one $\rm{U(1)}$ symmetry group, generated by $\sum_j(-1)^{j} {S_{z,j}}^2$, due to the constraints:
${S_{x,j}}^2+{S_{y,j}}^2+{S_{z,j}}^2=2$.

For $J_x>J_z>0$, there are two distinct choices for the highest weight state $|\rm{hws}\rangle$:
(i) $|{\rm hws}\rangle=|1_z\cdots 1_z\rangle$ and
(ii) $|{\rm hws}\rangle=|0_x0_y\cdots 0_x0_y\rangle$,
in the sense that the first choice is invariant under the one-site translation, whereas the second choice is not invariant under the one-site translation.
Here, $|1_z\rangle$ is the eigenvector of $S_{z,j}$, with the eigenvalue being $1$,
and $|0_x\rangle/|0_y\rangle$ is the eigenvector of $S_{x,j}/S_{y,j}$, with the eigenvalue being $0$.
However, the two choices are unitarily equivalent under a local unitary transformation $U$: $K_{x,j}\rightarrow K_{y,j}$, $K_{y,j}\rightarrow K_{z,j}$ and $K_{z,j}\rightarrow K_{x,j}$.
As a consequence, the entanglement entropy $S(n)$ for the degenerate ground states, corresponding to the two choices, must be identical.
Therefore, we only need to focus on the first choice for brevity. Note that the action of $K_{-,j}$ and $K_{+,j}$ on $|1_z\rangle_j$ takes the form: $K_{-,j}|1_z\rangle_j=(-1)^{j}\sqrt{2}/2|-1_z\rangle_j$ and $K_{+,j}|1_z\rangle_j=0$.

The interpolating fields are $K_{+,j}$ and $K_{-,j}$, for the generator $K_-$ and the generator $K_+$, respectively.
Thus, $\langle K_{z,j}\rangle$ is the local order parameter, given $\langle[K_{+,j},K_-]\rangle=\langle[K_+,K_{-,j}]\rangle=\langle K_{z,j}\rangle\neq0$.
Therefore, the two symmetry generators $K_-$ and $K_+$ are spontaneously broken.
According to the counting rule ~(\ref{nab}), there is one type-B GM.
In addition, since $K_{-,j}|1_z\rangle_j\neq0$, but $K_{-,j}^{\;\;\;2}|1_z\rangle_j=0$, we have $q=1$.
This is a specific realization of the general scheme: $K_+=E_1$, $K_-=F_1$ and $K_z=H_1$.

A sequence of degenerate ground states $|L,M\rangle$ are generated from the repeated action of the lowering operator $K_-$ on the highest weight state $|1_z\cdots 1_z\rangle$:
\begin{equation}
 |L,M\rangle=\frac{1}{Z(L,M)}K_-^{\,\,\,M}|1_z\cdots 1_z\rangle,
\end{equation}
with
\begin{equation}
Z(L,M)=M!\sqrt{\frac{C_L^M}{2^M}}.
\label{zlmcasei}
\end{equation}
We remark that $|L,M\rangle$ $(M=0,\ldots, L)$ span a $L+1$-dimensional irreducible representation of the symmetry group $\rm{SU(2)}$.
A derivation of the concrete expression for $Z(L,M)$ is presented in Sec. E of the SM.

The degenerate ground states $|L,M\rangle$ admit an exact Schmidt  decomposition:
\begin{equation}
|L,M\rangle=\sum_{k=0}^{\min{(M,n)}}\lambda(L,n,M,k)|n,k\rangle|L-n,M-k\rangle,
\label{svdcasei}
\end{equation}
where the Schmidt coefficients $\lambda(L,n,M,k)$ take the form,
\begin{equation}
\lambda(L,n,M,k)=\sqrt{\frac{C_n^kC_{L-n}^{M-k}}{C_L^M}}.
\label{casei}
\end{equation}
Therefore, the eigenvalues $\Lambda(L,n,M,k)$ of the reduced density matrix $\rho_L(n,M)$ are $\Lambda(L,n,M,k)=[\lambda(L,n,M,k)]^2$.
Hence, the entanglement entropy $S_L(n,M)$ follows from Eq.~(\ref{snk}).
An analytical treatment confirms the logarithmic scaling behaviour, as predicted in Eq.~(\ref{sn}) (cf. Sec. B of the SM).

\begin{figure}
  \centering
\includegraphics[height=4.5cm,width=0.33\textwidth]{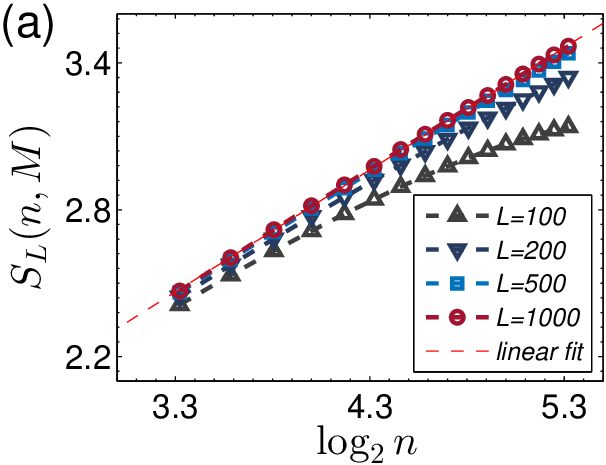}
\hspace{5mm}
\includegraphics[height=4.5cm,width=0.33\textwidth]{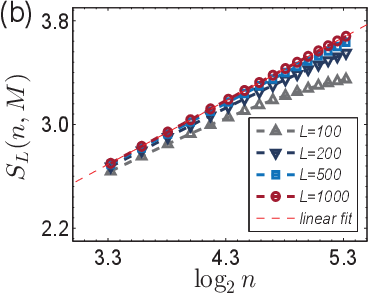}
\caption{The entanglement entropy $S_L(n,M)$ vs $\log_2 n$ for the coexisting fractal states: (a) $M=L/2$ and (b) $M=L/4$, when $L$ is varied: $L=100$, $200$, $500$, and $1000$.
A significant deviation from the logarithmic scaling behaviour is observed when $L$ is relatively small, but tends to vanish, as $L$ increases.
The prefactor is close to the exact value $1/2$, with an error being less than 2$\%$, when $L=1000$: $S_{1000}(n,250) = 0.499\log_2n + 0.818$ and $S_{1000}(n,500) = 0.490\log_2n + 1.080$ for $M=L/2$ and $M=L/4$, respectively.
Here, $n$ ranges from $10$ to $40$.}
\label{entropyphiyz}
\end{figure}

In order to understand how the logarithmic scaling behaviour emerges in the thermodynamic limit, we plot $S_L(n,M)$ vs $\log_2 n$ in Fig.~\ref{entropyphiyz} for the coexisting fractal states: (a) $M=L/2$ and (b) $M=L/4$, when $L$ is varied: $L=100$, $200$, $500$, and $1000$.
A significant deviation from the logarithmic scaling behaviour is observed when $L$ is relatively small, but tends to vanish, as $L$ increases.
The prefactor is close to the exact value $1/2$, with an error being less than 2$\%$, when $L=1000$: $S_{1000}(n,250) = 0.499\log_2n + 0.818$ and $S_{1000}(n,500) = 0.490\log_2n + 1.080$ for $M=L/2$ and $M=L/4$, respectively.
Here, $n$ ranges from $10$ to $40$.
Our numerics confirm that, for large enough $L$,  $S_L(n,M)$ logarithmically increases with $n$, as $n$ increases, subject to $1\ll n \ll L/2$, irrespective of $M$, as long as $M/L$ is neither 0 nor 1.  That is, for the block size $n$ between 10 and 40,  it is necessary to choose, e.g., $L=1000$,  to ensure that it is large enough to reach a scaling limit.

\section{summary}
In this work, a systematic investigation has been performed for  a set of orthonormal basis states, which appear as highly degenerate ground states arising from SSB with a type-B GM in a quantum many-body system from a quantum entanglement perspective. It is found that the orthonormal basis states admit an exact Schmidt decomposition, which unveils their self-similarity relevant to scale invariance. This implies that the entanglement entropy $S(n)$ scales logarithmically with the block size $n$ in the thermodynamic limit, with the prefactor being half the number of type-B GMs $N_B$, as far as the orthonormal basis states are concerned.
Meanwhile, as follows from a field-theoretic prediction~\cite{doyon}, the prefactor is half the fractal dimension $d_f$.
Therefore, the fractal dimension $d_f$ is identical to the number of type-B GMs $N_B$ for the orthonormal basis states arising from SSB with type-B GMs.
Indeed, this conclusion has now been further confirmed in a forthcoming article~\cite{cantorset}, where a distinction between an intrinsic abstract fractal underlying the ground state subspace, as anticipated here, and an extrinsic fractal introduced to reveal the intrinsic abstract fractal, has been made.
In addition, an extensive numerical analysis has been performed to reveal how the logarithmic scaling behaviour emerges, as the system size $L$ increases, for the $\rm{SU(2)}$ spin-$s$ ferromagnetic model, the $\rm{SU(N+1)}$ ferromagnetic model, with $N=2$, $3$ and $4$, and the $\rm{SU(2)}$ spin-1 anisotropic biquadratic model, with the fractal dimension $d_f$ being 1, $N$, and 1, respectively. Accordingly, the number of type-B GMs is 1, $N$, and 1, respectively. This lends further support to our claim.

In fact,  three independent approaches have been presented to justify our prediction (\ref{sn})  for the three models under investigation. The first is based on our extensive numerical calculations to demonstrate how such a logarithmic scaling relation emerges when the system size tends to infinity. The second is an analytical approach to
the entanglement entropy in the thermodynamic limit (cf. Sec. B of the SM). The third is based on a heuristic argument presented in Sec. A of the SM. They yield consistent results, thus producing compelling evidence for the validity of our prediction. We emphasize that, although the first two approaches are only applicable to the three models under investigation, the third approach should be valid for any quantum many-body systems undergoing SSB with type-B GMs. We remark that scale-invariant states arising from SSB with type-B GMs are not conformally invariant. Hence, the number of type-B GMs plays a similar role to  central charge in conformal field theory~\cite{vidal,cardy}, since both of them count the number of low-energy gapless excitations. In other words, there are two distinct types of scale-invariant states, which may be distinguished from their distinct universal finite-size scaling relations, as well as from distinct behaviors under different types of the boundary conditions, as argued in our subsequent work~\cite{finitesize}.

In closing, a few remarks are in order. First, the extension to a symmetry group $G$ other than $\rm{SU(N+1)}$ is possible, although our discussion focuses on $\rm{SU(N+1)}$.
Second, it is anticipated to extend our prediction to quantum many-body systems in two and higher spatial dimensions, modulo a few subtle issues~\cite{2d}, given that the occurrence of SSB with a type-B GM does not depend on the spatial dimensionality.
Last but not least, the scaling behaviour of the entanglement entropy remains unclear in the thermodynamic limit when both type-A and type-B GMs are present in a quantum many-body system.
It is anticipated that an extension to higher spatial dimensions raises a few conceptually interesting questions to be addressed, and the situation becomes much more complicated if type-A GMs are involved.

\section{Acknowledgements}
We thank Murray Batchelor, Sam Young Cho, and John Fjaerestad for enlightening discussions.
The work is supported by the National Natural Science Foundation of China (Grant No. 11805285).

\end{document}


\title{Supplementary Material for ``Fractal dimension and the counting rule of the Goldstone modes"}
	
\author{Qian-Qian Shi}
\affiliation{Centre for Modern Physics,
	Chongqing University, Chongqing 400044, The People's Republic of
	China}

\author{Yan-Wei Dai}
\affiliation{Centre for Modern Physics,
	Chongqing University, Chongqing 400044, The People's Republic of
	China}

\author{Huan-Qiang Zhou}\thanks{E-mail: hqzhou@cqu.edu.cn}
\affiliation{Centre for Modern Physics,
	Chongqing University, Chongqing 400044, The People's Republic of
	China}

\author{Ian P. McCulloch}
\affiliation{Department of Physics, National Tsing Hua University, Hsinchu 30013, Taiwan}
\affiliation{School of Mathematics and Physics, The University of Queensland, St. Lucia, QLD 4072, Australia}
\affiliation{Centre for Modern Physics, Chongqing University, Chongqing 400044, The People's Republic of China}

\maketitle

\subsection{A logarithmic scaling relation between the entanglement entropy $S(n)$ and the block size $n$}\label{scaleinvariant}

A specific set of  the orthonormal basis states, defined by Eq.(2) in the main text, span the ground state subspace, which admit an exact Schmidt decomposition (3).  That is, the block, the environment and the system share the same type of orthonormal basis states, indicating the self-similarity in the real space.  Actually, the three sets of the basis states for the block, the environment and the system itself are formally identical, since they are generated from the action of lowering operators (of the same group) on the highest weight state, defined for the block, the environment and the system, respectively.  In this sense, they are similar to each other, thus implying that the self-similarity in the real space manifests itself in the exact Schmidt decomposition. Hence, one may expect that highly degenerate ground states arising from SSB with type-B GMs are scale-invariant.

We present a heuristic argument, aiming to unveil a logarithmic scaling relation between the entanglement entropy $S(n)$ and the block size $n$, for a scale-invariant state, in the thermodynamic limit, when the fillings $f_1,\ldots,f_N$ are kept to be constant.
We focus on scale-invariant states in one spatial dimension, with the block $B$ consisting of $n$ contiguous lattice sites for simplicity. The partition of the system into the block $B$ and the environment $E$ amounts to introducing a length scale $n$, thus the system is expected to react.  Physically, it is legitimate to consider an effective (continuum) field theory, for a quantum many-body system on a lattice, in the long wavelength limit, when the thermodynamic limit is approached.

Suppose the entanglement entropy $S(n)$ for a specific (degenerate) ground state is $S(n)=f(n)$, with $f(n)$ being a function of $n$ to be determined. Generically, such a  degenerate ground state may be a linear combination of the orthonormal basis states in Eq.(2). The transformation $n \rightarrow \lambda \; n$ amounts to introducing two sequences of the values of the block size $n$, with a dimensionless ratio $\lambda$. The scale invariance implies that the scaling behaviour with $n$ must remain the same, with an additional contribution from $\lambda$ being additive. Mathematically, we have
\begin{equation}
f(\lambda \; n)=f(n)+F(\lambda),
\label{fnF}
\end{equation}
where $F(\lambda)$ is a function of $\lambda$ to be determined.
It is easy to see that
\begin{equation}
 F(1)=0.
\end{equation}
Taking the first-order derivative with respect to the parameter $\lambda$ on both hand sides of Eq.(\ref{fnF}), we have
\begin{equation}
nf'(\lambda\; n)=F'(\lambda).
\end{equation}
Setting $y=\lambda n$, we are led to
\begin{equation}
 yf'(y)=\lambda F'(\lambda)
 \end{equation}
The variables are separated, since the left hand side only depends on $y$, and the right hand side only depends on $\lambda$.  This implies that
\begin{equation}
yf'(y)=\kappa,
\end{equation}
and
\begin{equation}
\lambda F'(\lambda)=\kappa,
\end{equation}
with $\kappa$ being a constant to be determined.
This yields
\begin{equation}
f(y)=\kappa\ln{y}+f(1),
\label{fkapa}
\end{equation}
and
\begin{equation}
F(\lambda)=\kappa\ln{\lambda}.
\label{fkapa2}
\end{equation}
Now it comes to a crucial question concerning how the number of type-B GMs enters the prefactor $\kappa$ for this type of scale-invariant state.
If the fillings $f_1,\ldots,f_N$ are zero, then the ground state is the highest weight state, which is unentangled. Thus, the entanglement entropy $S(n)$ is simply zero. This implies that the entanglement entropy $S(n)$ features a singularity, with a discontinuous jump in the prefactor in front of the logarithmic function, when the zero filling limit is approached. Physically, this is due to the fact that the highest weight state, as an unentangled ground state, is special, in the sense that type-B GMs manifest themselves only when the other degenerate ground states with non-zero fillings are involved. Note that the same argument is valid for the lowest weight state, which is an unentangled ground state, when the fillings $f_1,\ldots,f_N$ reach the maximum values.
In other words, the fillings $f_1,\ldots,f_N$ play an important role. As a result, we have $\kappa =0$, indicating that none of type-B GMs is sensed for both the highest and lowest weight states. 

If the values of the fillings are fixed to be non-zero or the maximum, then one may expect that $\kappa$ should be linearly proportional to the number of type-B GMs: $\kappa=\eta N_B$, with $\eta$ being a universal constant.  In order to establish this fact, we remark that $\kappa$ must be a function of  the number of type-B GMs $N_B$, since only the low-lying excitations make a contribution to the scaling behaviour of the entanglement entropy $S(n)$ with the block size $n$, which is universal. That is, $\kappa = \kappa(N_B)$. To proceed, imagine a fictitious composite system consisting of two arbitrary (interacting) subsystems that are not coupled to each other, with the numbers of type-B GMs being $ N_{B,1}$ and $N_{B,2}$, respectively, as long as they are defined on the same lattice. Then, the total number of type-B GMs for the fictitious system is $N_B= N_{B,1}+N_{B,2}$. Accordingly, we have $\kappa(N_B)=\kappa( N_{B,1})+\kappa(N_{B,2})$, given that the entanglement entropy for the fictitious system is the sum of the counterparts for the two subsystems, since a (degenerate) ground state must be factorized into a tensor product of two (degenerate) ground states in the two subsystems. This implies that $\kappa(N_B)$ is linearly proportional to $N_B$.  Here we emphasize that the linear proportionality is imposed on both the (fictitious) composite system and the two subsystems, with the subsystems being any arbitrary quantum many-body systems undergoing SSB with type-B GMs. That is, the value of $\eta$ is identical to the composite system and the subsystems. Hence, $\eta$ is independent of the specifics of the subsystems, and is thus universal, given that the subsystems themselves represent any quantum many-body systems undergoing SSB with type-B GMs.  Indeed, the value of $\eta$ may be determined from a specific model, as long as its exact scaling relation between the entanglement entropy $S(n)$ and the block size $n$, together with the number of type-B GMs $N_B$, are known.  For the $\rm{SU(2)}$ spin-$1/2$ ferromagnetic states, an analytic treatment yields that $S(n)=1/2\ln n+S_0$~\cite{salerno}, with the number of type-B GMs $N_B=1$~\cite{watanabe}.
As a consequence, we have $\eta=1/2$ for  the orthonormal basis states generated from the action of the lowering operators on the highest weight state for a generic quantum many-body system undergoing SSB with type-B GMs. That is, for the orthonormal basis states, $S(n)$ scales logarithmically with $n$,

\begin{equation}
S(n)=\frac{N_B}{2}\ln n+S_0,
\label{snb}
\end{equation}
where $S_0$ is an additive contribution to the entanglement entropy, which is non-universal.
In addition, a field-theoretic approach~\cite{doyon} predicts that
\begin{equation}
S(n)=\frac{d_f}{2}\ln n +S_0,
 \label{sdf}
 \end{equation}
with $d_f$ being the fractal dimension. Here, we stress that this scaling relation is valid for any linear combination of the overcomplete (non-orthogonal) basis states as a subset of the coset space~\cite{doyon}. 
Therefore, we are led to the identification of the fractal dimension $d_f$ with the number of type-B GMs: $d_f=N_B$, for  the orthonormal basis states arising from SSB with a type-B GM (for more details, we refer to a forthcoming article~\cite{cantorset}, where an alternative way to establish the scaling relation (\ref{sdf}) is also presented).

The argument may be extended to a scale-invariant state in $D$ spatial dimensions.
Then, (\ref{snb}) and (\ref{sdf}) become
\begin{equation}
S(n)=\frac{DN_B}{2}\ln n+S_0,
\label{snbD}
\end{equation}
and
\begin{equation}
S(n)=\frac{Dd_f}{2}\ln n +S_0,
 \label{sdfD}
 \end{equation}
respectively.
Here, we stress that $n$ should be understood as the linear size of the $D$-dimensional block.

In passing, we remark that our argument for the logarithmic scaling behavior, up to Eq.~(\ref{fkapa}) and Eq.~(\ref{fkapa2}), also works for conformally invariant states in the thermodynamic limit, since scale invariance is part of conformal invariance.
In fact, the prefactor $\kappa$ must be proportional to central charge: $\kappa=\zeta c$, with $\zeta$ being a constant, given that $c$ counts the number of gapless excitations~\cite{cft}.
One may also determine the proportionality constant $\zeta$ to be $1/3$ from the XY model in a longitudinal magnetic field -- an exactly solvable model~\cite{Jin}, with central charge $c=1$ at criticality~\cite{qpt}.
This allows us to reproduce the scaling relation of the entanglement entropy $S(n)$ with the block size $n$ for conformally invariant states~\cite{vidal,cardy}.

\subsection{An analytical approach to the entanglement entropy in the thermodynamic limit: the three illustrative models}

We present an analytical approach to the entanglement entropy for the three illustrative  models when the system size $L$ tends to infinity. For brevity, we restrict ourselves to the fundamental representation of $\rm{SU(N+1)}$, because it is sufficient to describe the entanglement entropy for the three models, except for the $\rm{SU(2)}$ spin-$s$ ferromagnetic states when $s >1/2$. Indeed, for both the $\rm{SU(2)}$ spin-$1/2$ ferromagnetic states and  the coexisting fractal states, the Schmidt coefficients $\lambda_L(n,k,M)$  take the same form as that for the $\rm{SU(N+1)}$ ferromagnetic states in Eq.~(20) with $N=1$. Here, we remark that our analysis is essentially adapted from that for permutation-invariant quantum states~\cite{salerno}.
	
Note that at fixed fillings $f_\alpha=M_\alpha/L$ ($\alpha=1$,\ldots, $N$), the entanglement entropy $S_{\!L}(n, M_1,\ldots, M_N)$ becomes $S(n,f_1,\ldots, f_N)$ in the thermodynamic limit $L \rightarrow \infty$.
More precisely, $S(n,f_1,\ldots, f_N)$ is defined as follows
	\begin{widetext}
		\begin{align}
			S(n,f_1,\ldots,f_N)=-\sum\limits_{k_1,\ldots,k_N}^{n}
			\Lambda_\infty(n,k_1,\ldots,k_N, f_1,\ldots,f_N)
			\log_{2}\Lambda_\infty(n,k_1,\ldots,k_N, f_1,\ldots,f_N),
			\label{snkf1n}
		\end{align}
	\end{widetext}
	where the sum $\sum_{k_1,\ldots,k_N}^{n}$ is taken over all the possible values of $0,\ldots, n$, subject to the constraints: $\sum_{\alpha=1}^N k_\alpha\le n$, and $\Lambda_\infty(n,k_1,\ldots,k_N,f_1,\ldots,f_N)$ denotes
	the eigenvalues of the reduced density matrix $\Lambda(L,n,k_1,\ldots,k_N,M_1,\ldots,M_N)$, determined as the square of the Schmidt coefficients in Eq.~(20) in the thermodynamic limit $L \rightarrow \infty$. Indeed, using the Stirling's approximation $m!=m^m\exp(-m)\sqrt{2\pi m}$, it takes the form
	\begin{widetext}
		\begin{equation}
			\Lambda_\infty(n,k_1,\ldots,k_N,f_1,\ldots,f_N)=
			\prod_{\alpha=1}^N C_{n-\sum_{\beta=1}^{\alpha-1}k_\beta}^{k_\alpha}u_\alpha^{k_\alpha}(1-u_\alpha)^{n-\sum_{\beta=1}^{\alpha}k_\beta},
		\end{equation}
		where $u_\alpha=f_\alpha/(1-\sum_{\beta=1}^{\alpha-1}f_\beta)$.
		It may be rewritten as
		\begin{equation}
			\Lambda_\infty(n,k_1,\ldots,k_N,f_1,\ldots,f_N)= n!\frac{v^t}{t!}\prod_{\alpha=1}^{N}\frac{f_\alpha^{k_\alpha}}{k_\alpha!},\label{laminfty}
		\end{equation}
		with $v=1-\sum_{\alpha=1}^{N}f_\alpha$ and $t=n-\sum_{\alpha=1}^{N}k_\alpha$. Introducing $f_{N+1}=v$ and $k_{N+1}=t$, we have $\sum_{\alpha=1}^{N+1}f_\alpha=1$. Note that $(\sum_{\alpha=1}^{N+1}f_\alpha)^n=\sum_{k_1,\ldots,k_N}\Lambda_\infty(n,k_1,\ldots,k_N,f_1,\ldots,f_N)=1$. In fact,  the distribution of $\Lambda_\infty(n,k_1,\ldots,k_N,f_1,\ldots,f_N)$ may be recognized as the multinomial distribution of $k_{\alpha}$ ($\alpha=1$,\ldots, $N+1$). Hence, we are led to
		\begin{align}
			\langle k_\alpha\rangle=n f_\alpha,\label{pq0}\\
			\langle (k_\alpha-\langle k_\alpha\rangle^2)\rangle=n f_\alpha(1-f_\alpha),\label{pq1}\\
			\langle (k_\alpha-\langle k_\alpha\rangle)(k_\beta-\langle k_\beta\rangle)\rangle=-n f_\alpha f_\beta, \; \alpha\neq\beta.\label{pq2}
		\end{align}
 
	 	Define $x_\alpha=k_\alpha/n$ and denote $\Lambda_\infty(n,x_1,\ldots,x_N,f_1,\ldots,f_N)\equiv \Lambda_\infty(n,k_1,\ldots,k_N,f_1,\ldots,f_N)$, we may approximate the eigenvalues  $\Lambda_\infty(n,x_1,\ldots,x_N,f_1,\ldots,f_N)$ (cf. Eq.~(\ref{laminfty})) in terms of the multinomial distribution
		\begin{equation}
			\Lambda_\infty(n,x_1,\ldots,x_N,f_1,\ldots,f_N) \approx \sqrt{\frac{{\rm det} A}{(2\pi)^N}}\exp{(-\frac{1}{2}\sum_{\alpha,\beta=1}^{N}A_{\alpha,\beta}(x_\alpha-f_\alpha)(x_\beta-f_\beta))},
			\label{detAlam}
		\end{equation}
	\end{widetext}
	where $x_\alpha$ and $x_\beta$ are shifted to take into account that the mean in Eq.~(\ref{pq0}) is non-zero. 
	In addition, the moments from the distribution (cf. Eq.~(\ref{detAlam})) give
	\begin{equation}
		\langle(x_\alpha-f_\alpha)(x_\beta-f_\beta)\rangle=\frac{M_{\alpha\beta}}{{\rm det}A},
		\label{xalphabeta}
	\end{equation}
	where $M_{\alpha\beta}$ represent the minors of the matrix $A$.  Comparing Eq.~(\ref{xalphabeta}) with Eqs.~(\ref{pq1}) and ~(\ref{pq2}), we are able to determine the elements $A_{\alpha\beta}$ of the matrix $A$. 
	Specifically, according to Eqs.~(\ref{pq1}) and ~(\ref{pq2}), we have $A_{\alpha\beta}=A_{\beta\alpha}$. Therefore, there are in total $N(N+1)/2$  elements in the matrix $A$ that need to be fixed.
	
	Once this is done, the entanglement entropy may be evaluated from
	\begin{widetext}
		\begin{equation}
			S(n,f_1,\ldots,f_N)\approx-\int dy_1\cdots \int dy_{N+1}[\Lambda_\infty(n,y_1,y_2,...,y_{N+1})\log_2\Lambda_\infty(n,y_1,y_2,...,y_{N+1})],
		\end{equation}
	\end{widetext}
	where $\Lambda_\infty(n,y_1,y_2,...,y_{N+1})$ has been introduced to denote $\Lambda_\infty(n,x_1,\ldots,x_N,f_1,\ldots,f_N)$, with $y_\alpha=x_\alpha-f_\alpha$ ($\alpha=1$,\ldots, $N+1$).
	For finite $f_\alpha$, the dominant contribution to the integral originates from the neighbor of the origin $y_1=y_2=\cdots y_{N+1}=0$. Accordingly, we may extend the integration to the whole real axis and perform the integral exactly.  The entanglement entropy $S(n,f_1,\ldots,f_N)$ thus takes the form
	\begin{equation}
		S(n,f_1,\ldots,f_N)\approx\frac{N}{2}\log_{2}(2\pi e n)+C(f_1,\ldots,f_N),
	\end{equation}
	with $C(f_1,\ldots,f_N)=1/2\log_2[(1-\sum_{\alpha=1}^{N}f_\alpha)\prod_{k=1}^{N} f_\alpha]$.

	This confirms our argument that the prefactor in front of $\log_{2}n$ is half the number of type-B GMs $N_B$ for this type of scale-invariant states, as long as the fillings $f_\alpha$ are non-zero or the maxima.

\subsection{A derivation of $Z(L,M)$ for the $\rm{SU(2)}$ spin-$s$ ferromagnetic states}\label{normsu2}
For the $\rm{SU(2)}$ spin-$s$ ferromagnetic states $|L,M\rangle$ in Eq.\;(9), we need to figure out a way to derive the normalization factor $Z(L,M)$.
Here, we resort to the permutation invariance of $|L,M\rangle$ to facilitate the derivation.

For our purpose, it is convenient to introduce a set of permutation-invariant states $|\psi(N_{-s},\ldots,N_{s})\rangle$, defined as follows
\begin{equation}
|\psi(N_{-s},\ldots,N_{s})\rangle=\frac{1}{Z_\psi}
\sum_{P}{|\underbrace{-s\cdots-s}_{N_{-s}}\mid\cdots \mid\underbrace{s\cdots s}_{N_{s}}\rangle},
\label{psins}
\end{equation}
with
\begin{equation}
Z_\psi=\sqrt{\prod_{r=-s}^{s-1}{C_{L-\sum_{m=-s}^{r-1}N_m}^{N_r}}}.
\end{equation}
Here, $Z_\psi$ is introduced to ensure that $|\psi(N_{-s},\ldots,N_{s})\rangle$ is normalized, and the sum $\sum_P$ is taken over all the possible permutations $P$ for a given partition $\{ N_{-s},\ldots,N_{s} \}$.

\begin{widetext}
Now we are ready to expand $|L,M\rangle$ in terms of the basis states $|\psi(N_{-s},\ldots,N_{s})\rangle$:
\begin{align}
|L,M\rangle = \frac{1}{Z(L,M)}
{\sum}'_{N_{-s},\ldots, N_s}c(L,M,N_{-\!s},\ldots ,N_{s})|\psi(N_{-\!s},\ldots,N_{s})\rangle,
\label{psisu2}
\end{align}
where
\begin{align}
c(L,M,N_{-s},\ldots,N_{s})=Z_\psi\frac{M!}{\sqrt{2^M}}\prod_{r=-s}^{s-1}\left[\sqrt{\varepsilon(s,r)}\right]^{N_{r}},\label{c}
\end{align}
with
\begin{equation*}
\varepsilon(s,r)=\frac{\prod_{m=r+1}^{s}{(s+m)(s-m+1)}}{\prod_{m=r}^{s-1}(s-m)^2}.
\end{equation*}
Here, the sum ${\sum}'_{N_{-s},\ldots, N_s}$ is taken over all the possible values of $N_{-s},\ldots, N_s$, subject to the constraints: $\sum_{m=-s}^s N_m=L$ and $\sum_{m=-s}^{s}(s-m)N_m=M$.
Indeed, $Z(L,M)$ takes the form
\begin{equation}
Z(L,M)=\sqrt{{\sum}'_{N_{-s},\ldots, N_s}c(L,M,N_{-s},\ldots,N_{s})^2}.\label{z}
\end{equation}
Substituting Eq.(\ref{c}) into Eq.(\ref{z}), we have
\begin{align}
Z(L,M)=
\frac{M!}{\sqrt{2^M}} \sqrt{{\sum}'_{N_{-s},\ldots, N_s}\prod_{r=-s}^{s-1}\left[\varepsilon(s,r)\right]^{N_r}
{C_{L-\sum_{m=-s}^{r-1}N_m}^{N_r}}}.
\end{align}
\end{widetext}

\subsection{A derivation of $Z(L,M_1,...,M_N)$ for the $\rm{SU(N+1)}$ ferromagnetic states}\label{normsuN}
In order to derive the normalization factor $Z(L,M_1,\ldots,M_N)$ for the $\rm{SU(N+1)}$ ferromagnetic states $|L,M_1,\ldots,M_N\rangle$ in
Eq.(13), we need to take advantage of the permutation invariance of $|L,M_1,\ldots,M_N\rangle$.
	\begin{widetext}
In fact, one may rewrite $|L,M_1,\ldots,M_N\rangle$ as follows
\begin{align}
|L,M_1,\ldots,M_N\rangle= \frac{\prod_{\alpha=1}^N M_\alpha!}{Z(L,M_1,\ldots,M_N)}
\!\sum_{P}|\underbrace{1\cdots1}_{L-\sum_{\alpha=1}^N M_{\alpha}}\mid\underbrace{2,\cdots,2}_{M_{1}} \cdots \mid\underbrace{N+1\cdots N+1}_{M_{N}}\rangle,
\end{align}
where $|\beta\rangle$ ($\beta=1$, 2, \ldots, $N+1$) are defined as a $N+1$-dimensional vector, with the $\beta$-th entry being $1$, and the others being 0, and the sum $\sum_P$ is taken over all the permutations $P$ for a given partition $\{L-\sum_{\alpha=1}^N M_{\alpha},M_1,\ldots,M_N\}$.  As a consequence,
$Z(L,M_1,\ldots,M_N)$ takes the form
\begin{equation}
Z(L,M_1,\ldots,M_N)=\prod_{\alpha=1}^N M_\alpha!\sqrt{C_{L-\sum_{\beta=1}^{\alpha-1}{M_\beta}}^{M_{\alpha}}}.
\end{equation}

\subsection{A derivation of $Z(L,M)$ for the coexisting fractal states}\label{normcf}
For the coexisting fractal states $|L,M\rangle$ in Eq.\;(18), we need to introduce a unit cell consisting of two nearest-neighbor sites, due to the staggered nature of the symmetry group $\rm{SU(2)}$. Therefore, there are four possible configurations: $|1_z{1}_z\rangle$, $|1_z{-1}_z\rangle$, $|-1_z{1}_z\rangle$, and $|-1_z{-1}_z\rangle$ in a unit cell. Here, $|\pm 1_z\rangle$ are the eigenvectors of $S_{z,j}$, with the eigenvalues being $\pm 1$.

One may rewrite $|L,M\rangle$ as follows
\begin{align}
|L,M\rangle\!=\!\frac{M!}{\sqrt{2^M}Z(L,M)}
{\sum}'_{N_{mm},N_{mp}, N_{pm},N_{pp}}\sum_{P}(-1)^{N_{mm}+N_{mp}}|\underbrace{{-1}_z{-1}_z\cdots {-1}_z{-1}_z}_{N_{mm}}
\mid\underbrace{-1_z1_z\cdots -1_z1_z}_{N_{mp}}\mid\underbrace{1_z-1_z\cdots 1_z-1_z}_{N_{pm}}\mid\underbrace{1_z1_z\cdots 1_z1_z}_{N_{pp}}\rangle,
\end{align}
where the sum $\sum_{P}$ is taken over all the permutations $P$ for a given partition $\{N_{mm},N_{mp},N_{pm},N_{pp}\}$, and the sum ${\sum}'_{N_{mm},N_{mp}, N_{pm},N_{pp}}$ is taken over all the possible values of $N_{mm},N_{mp}, N_{pm},N_{pp}$, subject to the constraints:
 $2N_{mm}+N_{mp}+N_{pm}=M$ and $N_{mm}+N_{mp}+N_{pm}+N_{pp}=L/2$. Here, $N_{mm},N_{mp}, N_{pm}$, and $N_{pp}$ denote the numbers of the unit cells in the configurations $|-1_z-{1}_z\rangle$, $|-1_z1_z\rangle$, $|1_z-{1}_z\rangle$, and $|1_z{1}_z\rangle$, respectively.

Then, $Z(L,M)$ takes the form
\begin{align}
Z(L,M)=\frac{M!}{\sqrt{2^M}}
\sqrt{{\sum}'_{N_{mm},N_{mp}, N_{pm},N_{pp}}C_{L/2}^{N_{mm}}C_{L/2-N_{mm}}^{N_{pp}}C_{L/2-N_{mm}-N_{pp}}^{N_{pm}}},
\end{align}
which may be simplified as
\begin{equation}
Z(L,M)=M!\sqrt{\frac{C_L^M }{2^M}}.
\end{equation}
\end{widetext}

\subsection{Acknowledgements}

We thank Murray Batchelor, Sam Young Cho, and John Fjaerestad for enlightening discussions.
The work is supported by the National Natural Science Foundation of China (Grant No. 11805285).